\documentclass[preprint,12pt,sort&compress]{elsarticle}
\usepackage[utf8]{inputenc}
\usepackage{color}
\usepackage{graphicx}
\usepackage{placeins}
\usepackage[margin=1in]{geometry}
\usepackage{appendix}
\usepackage{amsmath, amssymb, bm}
\usepackage{multirow}
\usepackage{placeins}
\usepackage{xfrac}
\usepackage[colorlinks=true, linkcolor=blue, citecolor=blue]{hyperref}
\usepackage[nameinlink, capitalize]{cleveref}
\usepackage[font=footnotesize, labelfont=bf]{caption}
\usepackage{subcaption}
\usepackage[version=4]{mhchem}
\usepackage{lineno}
\usepackage{bm}
\usepackage{comment}
\usepackage[T1]{fontenc}
\usepackage{anyfontsize}

\usepackage{tablefootnote}
\usepackage{float}
\usepackage{algorithm}
\usepackage{algorithmicx}
\usepackage{algpseudocode}

\journal{Computational Materials Science}

\begin{document}

\begin{frontmatter}

\title{The contribution of nitrogen Frenkel-pair formation to the high-temperature heat capacity of uranium mononitride}

\author[ncsu,unsw]{Mohamed AbdulHameed}
\author[ncsu,inl]{Benjamin Beeler\corref{1}}
\cortext[1]{Corresponding author}
\ead{bwbeeler@ncsu.edu}

\address[ncsu]{Department of Nuclear Engineering, North Carolina State University, Raleigh, NC 27695, USA}
\address[inl]{Idaho National Laboratory, Idaho Falls, ID 83415, USA}
\address[unsw]{School of Mechanical and Manufacturing Engineering, UNSW Sydney, Sydney, NSW 2052, Australia}

\begin{abstract}

The high-temperature heat capacity of uranium mononitride (UN) remains uncertain due to conflicting measurements and models above $\sim$1700~K. To assess whether intrinsic defect formation contributes to the observed superlinear behavior of $C_P(T)$, we perform large-scale molecular dynamics simulations using two interatomic potentials to quantify nitrogen diffusion and Frenkel-pair populations from 1800--2600~K. Both models show increasing anion mobility, but the Tseplyaev potential yields substantially larger Frenkel concentrations, producing a defect heat-capacity contribution of up to $\sim$10~J/(mol-K). This defect-driven term is consistent with the curvature seen in historical correlations and recent \textit{ab initio} results, suggesting that nitrogen sublattice disorder provides a plausible intrinsic mechanism for the high-temperature heat capacity of UN.

\end{abstract}


\end{frontmatter}


\section{Introduction}

The specific heat, $C_P$, of uranium mononitride (UN) is not yet conclusively established, with experimental datasets and theoretical predictions offering conflicting descriptions above roughly 1700~K. The most widely used correlation, developed by Hayes \textit{et al.}~\cite{Hayes1990IV}, exhibits a strongly superlinear increase in $C_P(T)$ at elevated temperatures, approaching a nearly $T^{5}$ dependence~\cite{AbdulHameed2024}. This trend originates from the high–temperature enthalpy measurements of Conway and Flagella~\cite{Conway1969}, which extend to approximately 2600~K and remain the only measurements reaching such elevated temperatures. The only other available high–temperature measurements, i.e., those by Affortit~\cite{Affortit1970} that extend up to about 2300~K, report an almost linear temperature dependence. They also report higher $C_P(T)$ values than all other datasets in the temperature range of 1200--1900~K. Thus, Hayes \textit{et al.}~\cite{Hayes1990IV} decided to disregard Affortit's dataset and rely only on the measurements of Conway and Flagella to extract $C_P(T)$ at high temperatures. 

A central complication arises from the composition of the Conway-Flagella samples, which contained roughly 20~wt.\% UO$_2$. Because UO$_2$ exhibits a premelting (superionic) transition driven by oxygen Frenkel defects~\cite{Pavlov2017,Cooper2014}, any attempt to extract a UN-specific enthalpy from mixed-material calorimetry requires subtracting a defect-rich UO$_2$ contribution whose behavior changes rapidly with temperature \cite{Hayes1990IV}. While the specific heat capacity extracted from Conway-Flagella enthalpy measurements agrees with other datasets at moderate temperature \cite{Hayes1990IV}, it develops a strong upward curvature when extrapolated above $\sim$1700~K. Given the known defect thermodynamics of UO$_2$ \cite{Pavlov2017,Cooper2014}, this trend could plausibly reflect the UO$_2$ component rather than the intrinsic response of UN \cite{Galvin2023}.

Theoretical predictions also diverge. Early \textit{ab initio} molecular dynamics (AIMD) simulations reported an nearly perfectly linear $C_{P}(T)$~\cite{Kocevski2023}. Classical molecular dynamics (MD) using the angular-dependent potential (ADP) of Tseplyaev and Starikov~\cite{Kuksin2016,Tseplyaev2016} reproduced a steep increase in $C_P(T)$ consistent with the Hayes correlation~\cite{AbdulHameed2024}. The embedded atom method (EAM) potential of Kocevski \textit{et al.}~\cite{Kocevski2022II} also predicts a nearly linear $C_P(T)$ at high temperatures \cite{AbdulHameed2024}. Recent AIMD simulations combined with the disorded local moment (DLM) method predict a $T^{2}$-like increase in $C_P(T)$~\cite{AbdulHameed2025}. This result is significant because AIMD+DLM is a first-principles method that treats both lattice dynamics and magnetic disorder on equal footing. The fact that a high-accuracy \textit{ab initio} calculation yields a moderate, $T^{2}$-like increase suggests that while the anomaly is likely less dramatic than the historical correlation implies, there is independent evidence that a nonlinear increase in $C_{P}(T)$ is an intrinsic feature of UN at high temperatures.

The unresolved question is not only the magnitude of the heat capacity but the physical origin of its temperature dependence. Galvin \textit{et al.} \cite{Galvin2023} argued that current data support two plausible interpretations: either the Kocevski potential correctly predicts an asymptotically linear high-temperature heat capacity and the Conway-Flagella curvature is extrinsic, likely influenced by UO$_2$ in the samples, or the stronger rise seen with the Tseplyaev potential reflects an intrinsic mechanism in UN that has not yet been experimentally confirmed. They also suggested that strong anharmonicity, phonon softening, or even partial UN decomposition could contribute to the observed rise in $C_P(T)$ if defect mechanisms alone are insufficient.

A useful reference is the defect thermodynamics of UO$_2$, where oxygen Frenkel pairs proliferate as the temperature approaches the superionic transition, producing large excess contributions to both enthalpy and heat capacity~\cite{Pavlov2017}. MD simulations by Cooper \textit{et al.}~\cite{Cooper2014} demonstrated that oxygen diffusion increases sharply and $C_{P}(T)$ exhibits a broad anomaly near 2700~K as the anion sublattice disorders. Because Conway–Flagella's samples contained enough UO$_2$ to contribute significantly to the measured enthalpy, it is plausible that the reconstructed UN correlation inherited part of this defect-driven curvature. However, the nonlinear $T^{2}$-like behavior obtained in recent AIMD+DLM simulations~\cite{AbdulHameed2025} suggests that a degree of intrinsic anion–sublattice softening may also occur in UN. This raises the question whether nitrogen Frenkel-pair formation in UN could contribute to UN's $C_P(T)$ analogous to that observed in UO$_2$.

\section{Computational details}

To address this question, we performed classical MD simulations of UN using two interatomic potentials: the Tseplyaev angular-dependent potential \cite{Tseplyaev2016} and the Kocevski EAM potential. All calculations eutilize the Large-scale Atomic/Molecular Massively Parallel Simulator (LAMMPS) software package \cite{Plimpton1995, Thompson2022} using a 1 fs time step, with periodic boundary conditions (PBCs) applied in all directions. The OVITO software package \cite{Stukowski2010} is used for supercell visualization and analysis. All simulations employ a $50 \times 50 \times 50$ supercell ($10^6$ atoms), allowing us to sample both transport and point-defect statistics over the  temperature range of 1800--2600~K. For diffusivity calculations, the supercells are equilibrated in the \textit{NPT} ensemble and then evolved for a total of 5~ns. The self-diffusion coefficients are obtained from the long-time limit of the mean-squared displacement, MSD,
\begin{equation}
D = \lim_{t\rightarrow\infty}\frac{\langle \Delta r^{2}(t) \rangle}{6t},
\end{equation}
using the slope over the final 3~ns of each trajectory. No defects were inserted in any of these simulations, and all contributions to the MSD originate from spontaneously produced Frenkel pairs. In all cases, uranium remained effectively immobile on these timescales, while nitrogen exhibited increasingly rapid motion with temperature. We note that these are transient, dynamically formed vacancy–interstitial pairs identified from instantaneous snapshots rather than static, equilibrium Frenkel defects in the classical thermodynamic sense; the term ``Frenkel pair'' is used throughout for convenience. The the goal of this direct MD-based defect count is to estimate the defect contribution to $C_P(T)$ in UN without \textit{a priori} assumptions about defect equilibrium or the functional form of defect concentration.

\section{Results}

\begin{figure*}[h!]
\centering
\begin{subfigure}{0.45\textwidth}
\includegraphics[width=\textwidth]{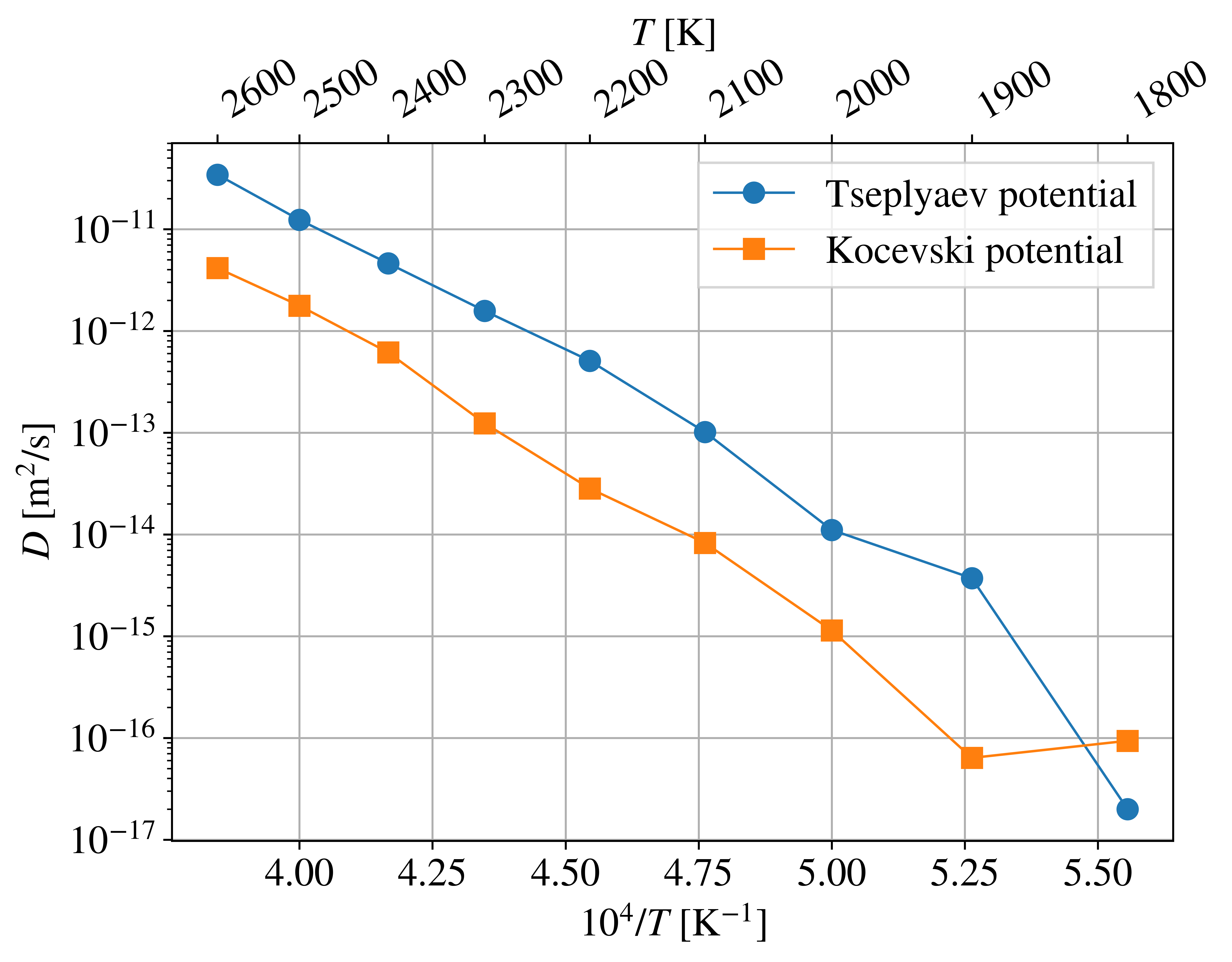}
\caption{}
\label{Fig:D}
\end{subfigure}
\hfill
\begin{subfigure}{0.45\textwidth}
\includegraphics[width=\textwidth]{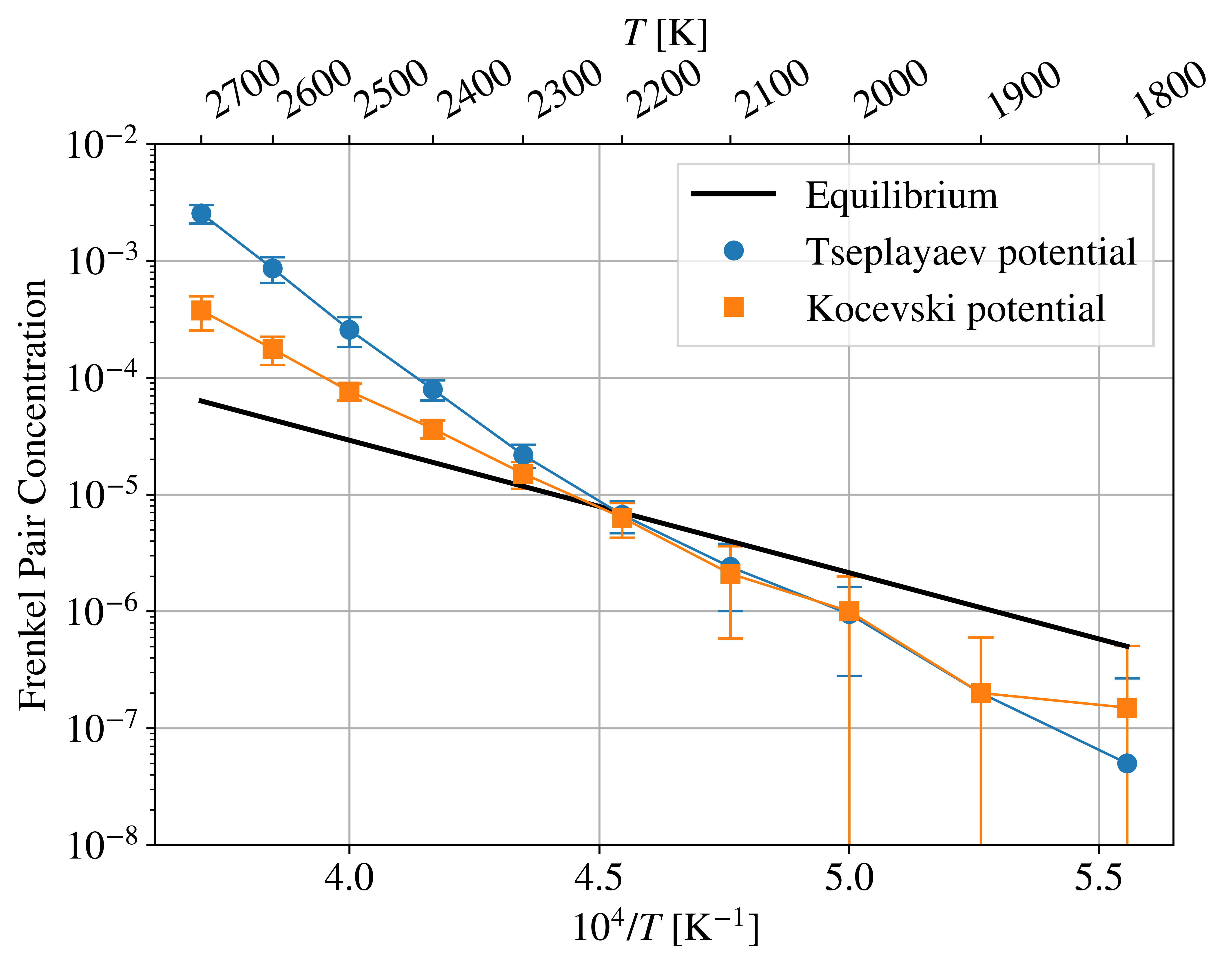}
\caption{}
\label{Fig:C}
\end{subfigure}
\caption{(Color online) (\textbf{a}) Nitrogen self-diffusion coefficients obtained from MSD analysis. (\textbf{b}) Nitrogen Frenkel-pair concentrations extracted using Wigner–Seitz defect analysis over 200~ps trajectories. Data points denote temporal averages over the post-equilibration sampling window, and error bars represent the standard deviation obtained from time averaging. The equilibrium concentration is based on Frenkel-pair formation energy of 4.5 eV calculated elsewhere using DFT and 0 K MD \cite{Yang2021,Kocevski2022I,AbdulHameed2024}.}
\end{figure*}

\cref{Fig:D} summarizes the nitrogen self-diffusion coefficients for both potentials. For the Tseplyaev potential, the nitrogen diffusivity, $D_\text{N}$, is essentially zero at temperatures below 1800~K and then increases sharply, rising from $3.7\times10^{-15}$~m$^{2}$/s at 1900~K to $3.4\times10^{-11}$~m$^{2}$/s at 2600~K, spanning nearly four orders of magnitude. The Kocevski potential yields the same qualitative trend but with systematically smaller values, typically one order of magnitude lower across the entire range, with $D_\text{N}$ approaching $4.1 \times 10^{-12}$~m$^{2}$/s at 2600~K. The increasing nitrogen transport above 1800--1900~K, especially for the Tseplyaev potential, coincides with the temperature range where the Conway-Flagella enthalpy data start deviating from an approximately linear trend into a pronounced nonlinear increase. This suggests that the same microscopic process responsible for the rapid growth of nitrogen mobility---namely, the onset of significant anion-sublattice disorder and Frenkel-pair formation---may also underlie the excess enthalpy that produces the apparent high-temperature anomaly in the empirical heat capacity.

To more directly quantify point-defect formation, we analyzed nitrogen Frenkel-pair concentrations using the Wigner-Seitz (WS) defect analysis implemented in OVITO \cite{Stukowski2010}. In this method, a defect-free reference lattice defines the positions of ideal sites, each associated with a Wigner-Seitz cell, i.e., the region of space closer to that site than to any other. For a given instantaneous configuration, each atom is assigned to the nearest reference site. Sites with occupancy zero correspond to vacancies, while sites with occupancy greater than one indicate interstitial defects. This nearest-site mapping provides a robust, topology-based identification of point defects. For each temperature, we simulated for 200~ps, recorded snapshots every 10~ps, and applied the WS analysis to all configurations after the first 10~ps to ensure equilibration. The reported Frenkel-pair concentrations correspond to the time-averaged mean over the sampled configurations, and the associated error bars represent the standard deviation over the temporal sampling window.

The resulting nitrogen Frenkel-pair concentrations, $c_{\mathrm{FP}}(T)$, are shown in \cref{Fig:C}. For the Tseplyaev potential, $c_{\mathrm{FP}}$ increases from $5\times10^{-8}$ at 1800~K to $8.6\times10^{-4}$ at 2600~K. Fitting an Arrhenius form,
\begin{equation}
c_{\mathrm{FP}}(T) = A \exp\!\left[-\frac{Q}{k_{\mathrm{B}}T}\right],
\end{equation}
yields an effective activation energy $Q_{\mathrm{ADP}} = 4.97$~eV, while the Kocevski potential produces a lower activation energy, $Q_{\mathrm{EAM}} = 3.79$~eV, and lower defect populations at all temperatures. Thus, the Tseplyaev potential generates nitrogen Frenkel concentrations large enough to potentially alter the thermodynamics.


Frenkel-pair formation contributes an additional term to the heat capacity \cite{Szwarc1969,Pavlov2017}:
\begin{equation}
C_{\mathrm{def}}(T) = \frac{1}{n_{\mathrm{moles}}} \frac{\partial}{\partial T} \left[n_{\mathrm{FP}}(T) \ Q_\text{FP} \right]
\approx \frac{Q_\text{FP}}{n_{\mathrm{moles}}} \frac{\partial n_{\mathrm{FP}}}{\partial T}.
\label{Eq:C_def}
\end{equation}
In practice, we evaluate $C_{\mathrm{def}}(T)$ directly from the Frenkel-pair populations obtained by the WS analysis for both interatomic potentials. We assume a formation enthalpy $Q$ = 4.5~eV, consistent with both the DFT estimates \cite{Yang2021,Kocevski2022I} and the $T \rightarrow 0$~K Frenkel pair formation energies of both the Tseplyaev and Kocevski potentials~\cite{AbdulHameed2024}. The temperature derivative in \cref{Eq:C_def} is computed using centered finite differences applied to the simulated defect populations.

\begin{figure*}[h!]
 \centering
 \includegraphics[width=0.5\textwidth]{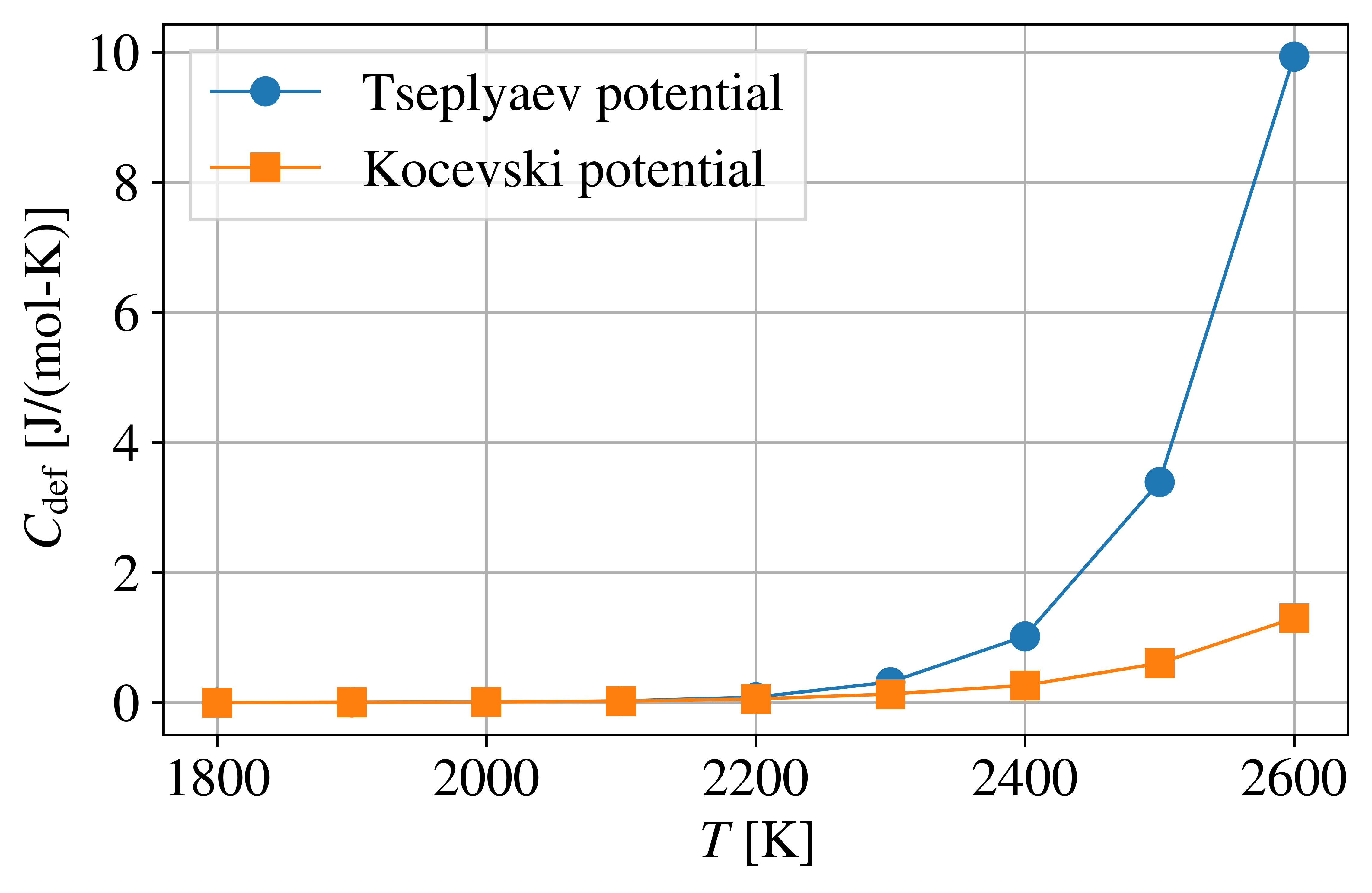}
 \caption{Defect-induced heat-capacity contribution $C_{\mathrm{def}}(T)$ computed from Wigner-Seitz Frenkel-pair statistics using both the Tseplyaev and Kocevski interatomic potentials.}
 \label{Fig:C_def}
\end{figure*}

The resulting $C_{\mathrm{def}}(T)$ curves are shown in \cref{Fig:C_def}. For the Tseplyaev potential, $C_{\mathrm{def}}$ reaches $\sim$10~J/(mol-K) at 2600~K, whereas for the Kocevski potential it remains on the order of 1~J/(mol-K) at the same temperature. Thus, the order-of-magnitude difference in Frenkel-pair concentration between the two potentials translates into an order-of-magnitude difference in the defect contribution to the heat capacity. This might provide an explanation for why the Tseplyaev potential reproduces the strong curvature of the Hayes correlation, while the Kocevski potential yields a nearly linear $C_{P}(T)$: the latter underestimates the magnitude of the Frenkel-pair contribution and therefore fails to reproduce the superlinear component in $C_P(T)$.

A key observation is that the temperature range in which both the nitrogen diffusivity and Frenkel-pair concentration increase most rapidly (roughly 1800--2000~K) coincides with the onset of deviations from a purely phononic heat capacity in experiment and AIMD+DLM. This parallel behavior strongly suggests that UN undergoes a form of anion-sublattice disordering crossover at high temperature, qualitatively analogous to the oxygen superionic transition in UO$_2$.


Our results therefore support a defect-driven interpretation of the high-temperature heat capacity of UN: nitrogen Frenkel-pair formation and associated sublattice disorder provide a plausible intrinsic mechanism for the observed superlinear $C_P(T)$. The calculated $C_{\mathrm{def}}(T)$ should be regarded as qualitative, since the defect populations are sensitive to the choice of interatomic potential and \cref{Eq:C_def} is a simplified estimate. Nevertheless, the order-of-magnitude difference in Frenkel-pair concentration between the two potentials produces a correspondingly large difference in the defect contribution to the heat capacity. Thus, while the Tseplyaev potential predicts defect populations large enough to reproduce the curvature in the Hayes correlation and the Kocevski potential yields a much weaker response, the true intrinsic behavior of UN potentially lies between these limits.

This hypothesis is experimentally testable. High-purity UN samples with carefully controlled oxygen content, combined with high-temperature calorimetry and tracer-diffusion measurements, are therefore essential to determine whether an anion-mobility crossover and correlated superlinear enhancement in $C_P(T)$ occur near the temperatures predicted here.

\section{Conclusions}

In this work, we used MD simulations with two interatomic potentials to investigate whether intrinsic nitrogen Frenkel-pair formation contributes to the high-temperature heat capacity of UN. Both potentials predict increasing nitrogen Frenkel-pair concentrations and diffusivities above 1800~K, but the Tseplyaev potential yields defect populations roughly an order of magnitude larger than the Kocevski potential across the entire temperature range. This difference translates directly into the thermodynamic response: $C_\text{def}$ reaches $\sim$10~J/(mol$\cdot$K) for the Tseplyaev potential at 2600~K but remains near $\sim$1~J/(mol$\cdot$K) for the Kocevski potential.

These results offer an atomistic explanation for the divergent heat-capacity predictions of the two potentials. When Frenkel pairs form in appreciable concentrations, the associated defect enthalpy produces a superlinear contribution to $C_P(T)$; when defect populations remain small, the heat capacity is dominated by lattice vibrations and remains nearly linear. This behavior resembles the defect-driven anomalies in UO$_2$, suggesting that UN may undergo a similar anion-sublattice disordering crossover at high temperature. Resolving the quantitative magnitude of this effect will require high-temperature calorimetry and tracer-diffusion measurements on high-purity UN samples.

\section{Discussion}

The defect contribution to the heat capacity estimated here via \cref{Eq:C_def} is simplified relative to the comprehensive statistical-thermodynamic treatment developed by Pavlov \textit{et al.} for UO$_2$~\cite{Pavlov2017}. Several approximations distinguish the present approach: (\textit{i}) the Frenkel-pair formation enthalpy $Q_\text{FP}$ is taken as temperature-independent, whereas Pavlov \textit{et al.} allowed it to decrease with temperature through Debye--H\"{u}ckel screening of charged defect interactions; (\textit{ii}) no non-configurational formation entropy is included; (\textit{iii}) defect populations are extracted directly from the MD trajectories rather than solved self-consistently from coupled mass-action equations; and (\textit{iv}) only a single defect type---nitrogen Frenkel pairs---is considered, neglecting, e.g., Schottky defects. Two physical features of UN, however, make these simplifications considerably more defensible than they would be for UO$_2$.

First, nitrogen Frenkel pairs in UN are electrically neutral because UN behaves as a metal~\cite{AbdulHameed2024b}, in contrast to the charged oxygen vacancy-interstitial pairs ($v_\text{O}^{\bullet\bullet}$ and $\text{O}_i^{\prime\prime}$) in UO$_2$. In UO$_2$, the net attractive force between oppositely charged defects reduces the effective formation enthalpy as their concentration increases, which in turn accelerates further defect production until the available interstitial sites begin to saturate, producing a maximum in $C_P(T)$~\cite{Pavlov2017,Cooper2014}. No such self-reinforcing mechanism operates for neutral defects, so treating $Q_\text{FP}$ as temperature-independent is appropriate for UN, and $C_\text{def}(T)$ is expected to increase monotonically until melting rather than peak, consistent with the behavior predicted by both potentials in this work.

Second, the omission of the non-configurational formation entropy can be justified quantitatively. Following the thermodynamic framework of Varotsos~\cite{Varotsos1976}, the formation entropy of a Frenkel defect can be written as:
\begin{equation}
S_\text{FP} = -\frac{Q_\text{FP}}{B_0}\frac{\partial B}{\partial T}.
\label{Eq:S_F}
\end{equation}
For a bulk modulus of the linear form $B(T) = B_0(1 - \alpha T)$, this simplifies to:
\begin{equation}
S_\text{FP} = Q_\text{FP} \ \alpha,
\end{equation}
independent of $B_0$. Using the temperature coefficient $\alpha = 2.375\times10^{-5}$~K$^{-1}$ for UN~\cite{Hayes1990II} and $Q_\text{FP} = 4.5$~eV gives $S_\text{FP} \approx 1.07\times10^{-4}$~eV/K~$\approx 1.24 \ k_B$. The corresponding Boltzmann prefactor is $\exp(S_\text{FP}/(2k_B)) \approx 1.86$, meaning that including entropy would uniformly scale the equilibrium Frenkel-pair concentration by a factor of approximately 1.86 at all temperatures, which is negligible relative to the order-of-magnitude difference in $c_\text{FP}$ between the two interatomic potentials. By comparison, Pavlov \textit{et al.}~\cite{Pavlov2017} used a formation entropy of $S_\text{FP} \approx 8.5 \ k_B$ for oxygen Frenkel pairs in UO$_2$, yielding $\exp(S_\text{FP}/(2k_B)) \approx 69.11$, roughly two orders of magnitude larger than the UN value.

\section*{Acknowledgments}

The authors thank Jacob Eapen for the fruitful discussions and useful suggestions. This research made use of the resources of the High-Performance Computing Center at Idaho National Laboratory, which is supported by the Office of Nuclear Energy of the U.S. Department of Energy and the Nuclear Science User Facilities under Contract No. DE-AC07-05ID14517.

\bibliographystyle{elsarticle-num}
\bibliography{ref}

\begin{thebibliography}{10}
\expandafter\ifx\csname url\endcsname\relax
  \def\url#1{\texttt{#1}}\fi
\expandafter\ifx\csname urlprefix\endcsname\relax\def\urlprefix{URL }\fi
\expandafter\ifx\csname href\endcsname\relax
  \def\href#1#2{#2} \def\path#1{#1}\fi

\bibitem{Hayes1990IV}
S.~L. Hayes, J.~K. Thomas, K.~L. Peddicord, Material property correlations for uranium mononitride {IV}. thermodynamic properties, Journal of Nuclear Materials 171 (1990) 300--318.

\bibitem{AbdulHameed2024}
M.~AbdulHameed, B.~Beeler, C.~O. Galvin, M.~W. Cooper, Assessment of uranium nitride interatomic potentials, Journal of Nuclear Materials 600 (2024) 155247.
\newblock \href {https://doi.org/https://doi.org/10.1016/j.jnucmat.2024.155247} {\path{doi:https://doi.org/10.1016/j.jnucmat.2024.155247}}.

\bibitem{Conway1969}
J.~B. Conway, P.~N. Flagella, Physical and mechanical properties of reactor materials, Tech. Rep. GEMP-1004, General Electric Company (10 1969).
\newblock \href {https://doi.org/10.2172/4815902} {\path{doi:10.2172/4815902}}.

\bibitem{Affortit1970}
C.~Affortit, Chaleur specifique de {UC} et {UN}, Journal of Nuclear Materials 34~(1) (1970) 105--107.
\newblock \href {https://doi.org/https://doi.org/10.1016/0022-3115(70)90014-0} {\path{doi:https://doi.org/10.1016/0022-3115(70)90014-0}}.

\bibitem{Pavlov2017}
T.~R. Pavlov, M.~R. Wenman, L.~Vlahovic, D.~Robba, R.~J. Konings, P.~V. Uffelen, R.~W. Grimes, Measurement and interpretation of the thermo-physical properties of {UO$_2$} at high temperatures: The viral effect of oxygen defects, Acta Materialia 139 (2017) 138--154.
\newblock \href {https://doi.org/10.1016/j.actamat.2017.07.060} {\path{doi:10.1016/j.actamat.2017.07.060}}.

\bibitem{Cooper2014}
M.~W. Cooper, S.~T. Murphy, P.~C. Fossati, M.~J. Rushton, R.~W. Grimes, Thermophysical and anion diffusion properties of ({U$_x$,Th$_{1-x}$)O$_2$}, Proceedings of the Royal Society A: Mathematical, Physical and Engineering Sciences 470 (2014).
\newblock \href {https://doi.org/10.1098/rspa.2014.0427} {\path{doi:10.1098/rspa.2014.0427}}.

\bibitem{Galvin2023}
C.~Galvin, N.~Kuganathana, N.~Barron, R.~Grimes, Predicted thermophysical properties of {UN}, {PuN}, and {(U, Pu)N}, Journal of Applied Physics 135 (2024).
\newblock \href {https://doi.org/10.1063/5.0177315} {\path{doi:10.1063/5.0177315}}.

\bibitem{Kocevski2023}
V.~Kocevski, D.~A. Rehn, A.~J. Terricabras, A.~van Veelen, M.~W. Cooper, S.~W. Paisner, S.~C. Vogel, J.~T. White, D.~A. Andersson, Finite temperature properties of uranium mononitride, Journal of Nuclear Materials 576 (2023) 154241.
\newblock \href {https://doi.org/10.1016/j.jnucmat.2023.154241} {\path{doi:10.1016/j.jnucmat.2023.154241}}.

\bibitem{Kuksin2016}
A.~Y. Kuksin, S.~V. Starikov, D.~E. Smirnova, V.~I. Tseplyaev, The diffusion of point defects in uranium mononitride: Combination of {DFT} and atomistic simulation with novel potential, Journal of Alloys and Compounds 658 (2016) 385--394.
\newblock \href {https://doi.org/10.1016/j.jallcom.2015.10.223} {\path{doi:10.1016/j.jallcom.2015.10.223}}.

\bibitem{Tseplyaev2016}
V.~I. Tseplyaev, S.~V. Starikov, The atomistic simulation of pressure-induced phase transition in uranium mononitride, Journal of Nuclear Materials 480 (2016) 7--14.
\newblock \href {https://doi.org/10.1016/j.jnucmat.2016.07.048} {\path{doi:10.1016/j.jnucmat.2016.07.048}}.

\bibitem{Kocevski2022II}
V.~Kocevski, M.~W. Cooper, A.~J. Claisse, D.~A. Andersson, Development and application of a uranium mononitride ({UN}) potential: Thermomechanical properties and {Xe} diffusion, Journal of Nuclear Materials 562 (2022).
\newblock \href {https://doi.org/10.1016/j.jnucmat.2022.153553} {\path{doi:10.1016/j.jnucmat.2022.153553}}.

\bibitem{AbdulHameed2025}
M.~AbdulHameed, B.~Beeler, A.~Claisse, Ab initio molecular dynamics of paramagnetic uranium mononitride ({UN}) using disordered local moments, Computational Materials Science 260 (2025) 114115.
\newblock \href {https://doi.org/https://doi.org/10.1016/j.commatsci.2025.114115} {\path{doi:https://doi.org/10.1016/j.commatsci.2025.114115}}.

\bibitem{Plimpton1995}
S.~Plimpton, Fast parallel algorithms for short-range molecular dynamics, Journal of Computational Physics 117 (1995) 1--19.
\newblock \href {https://doi.org/10.1006/JCPH.1995.1039} {\path{doi:10.1006/JCPH.1995.1039}}.

\bibitem{Thompson2022}
A.~P. Thompson, H.~M. Aktulga, R.~Berger, D.~S. Bolintineanu, W.~M. Brown, P.~S. Crozier, P.~J. in~'t Veld, A.~Kohlmeyer, S.~G. Moore, T.~D. Nguyen, R.~Shan, M.~J. Stevens, J.~Tranchida, C.~Trott, S.~J. Plimpton, {LAMMPS} - a flexible simulation tool for particle-based materials modeling at the atomic, meso, and continuum scales, Computer Physics Communications 271 (2022).
\newblock \href {https://doi.org/10.1016/j.cpc.2021.108171} {\path{doi:10.1016/j.cpc.2021.108171}}.

\bibitem{Stukowski2010}
A.~Stukowski, Visualization and analysis of atomistic simulation data with {OVITO}--the open visualization tool, Modelling and Simulation in Materials Science and Engineering 18 (2010).
\newblock \href {https://doi.org/10.1088/0965-0393/18/1/015012} {\path{doi:10.1088/0965-0393/18/1/015012}}.

\bibitem{Yang2021}
L.~Yang, N.~Kaltsoyannis, Incorporation of {Kr} and {Xe} in uranium mononitride: A density functional theory study, Journal of Physical Chemistry C 125 (2021) 26999--27008.
\newblock \href {https://doi.org/10.1021/acs.jpcc.1c08523} {\path{doi:10.1021/acs.jpcc.1c08523}}.

\bibitem{Kocevski2022I}
V.~Kocevski, D.~A. Rehn, M.~W.~D. Cooper, D.~A. Andersson, First-principles investigation of uranium mononitride ({UN}): Effect of magnetic ordering, spin-orbit interactions and exchange correlation functional, Journal of Nuclear Materials 559 (2022).
\newblock \href {https://doi.org/10.1016/j.jnucmat.2021.153401} {\path{doi:10.1016/j.jnucmat.2021.153401}}.

\bibitem{Szwarc1969}
R.~Szwarc, The defect contribution to the excess enthalpy of uranium dioxide: Calculation of the {Frenkel} energy, Journal of Physics and Chemistry of Solids 30 (1969).
\newblock \href {https://doi.org/10.1016/0022-3697(69)90024-9} {\path{doi:10.1016/0022-3697(69)90024-9}}.

\bibitem{AbdulHameed2024b}
M.~AbdulHameed, B.~Beeler, A.~Claisse, Atomistic investigation of plastic deformation and dislocation motion in uranium mononitride, Applied Sciences 15 (2024) 2666.
\newblock \href {https://doi.org/https://doi.org/10.3390/app15052666} {\path{doi:https://doi.org/10.3390/app15052666}}.

\bibitem{Varotsos1976}
P.~Varotsos, Comments on the formation entropy of a {Frenkel} defect in {BaF}$_2$ and {CaF}$_2$, Phys. Rev. B 13~(2) (1976) 938.
\newblock \href {https://doi.org/10.1103/PhysRevB.13.938} {\path{doi:10.1103/PhysRevB.13.938}}.

\bibitem{Hayes1990II}
S.~L. Hayes, J.~K. Thomas, K.~L. Peddicord, Material property correlations for uranium mononitride {II}. mechanical properties, Journal of Nuclear Materials 171 (1990) 271--288.

\end{thebibliography}

\end{document}